\documentclass{article}

\usepackage[final, nonatbib]{nips}

\usepackage[utf8]{inputenc} 
\usepackage[T1]{fontenc}    
\usepackage{url}            
\usepackage{booktabs}       
\usepackage{amsfonts}       
\usepackage{nicefrac}       
\usepackage{microtype}      

\usepackage{times}
\usepackage{amssymb}
\usepackage{subfigure}
\usepackage{graphicx}
\usepackage[colorlinks=true, allcolors=blue]{hyperref}
\usepackage{authblk}
\usepackage{makecell}
\usepackage{fixfoot}

\renewcommand{\Authands}{ and } 

\title{Hello Edge: Keyword Spotting on Microcontrollers}

\author[1,2]{Yundong Zhang}
\author[1]{Naveen Suda}
\author[1]{Liangzhen Lai}
\author[1]{Vikas Chandra}
\affil[1]{Arm, San Jose, CA}
\affil[2]{Stanford University, Stanford, CA}

\begin{document}
\maketitle

\begin{abstract}
Keyword spotting (KWS) is a critical component for enabling speech based user
interactions on smart devices. It requires real-time response and high accuracy
for good user experience.
Recently, neural networks have become an attractive choice for KWS architecture
because of their superior accuracy compared to traditional speech processing
algorithms.
Due to its always-on nature, KWS application has highly constrained power 
budget and typically runs on tiny microcontrollers with limited memory
and compute capability.
The design of neural network architecture for KWS must consider these
constraints.
In this work, we perform neural network architecture evaluation and exploration
for running KWS on resource-constrained microcontrollers.
We train various neural network architectures for keyword spotting
published in literature to compare their accuracy and 
memory/compute requirements.
We show that it is possible to optimize these neural network architectures to fit
within the memory and compute constraints of microcontrollers without
sacrificing accuracy.
We further explore the depthwise separable convolutional neural network (DS-CNN)
and compare it against other neural network architectures. 
DS-CNN achieves an accuracy of 95.4\%,
which is \textasciitilde10\% higher than the DNN model with similar
number of parameters. 

\end{abstract}

\let\thefootnote\relax\footnotetext{$^2$Work was done while the author was an intern at Arm.}

\section{Introduction}
Deep learning algorithms have evolved to a stage where they have surpassed human 
accuracies in a variety of cognitive tasks including image classification
~\cite{image_stateofart} and conversational speech recognition
~\cite{speech_stateofart}. 
Motivated by the recent breakthroughs in deep learning based speech recognition 
technologies, speech is increasingly becoming a more natural way to interact 
with consumer electronic devices, for example, Amazon Echo, Google Home and 
smart phones. However, always-on speech recognition is not 
energy-efficient and may also cause network congestion to transmit continuous
 audio stream from billions of these devices to the cloud. Furthermore, such 
a cloud based solution adds latency to the application, which hurts user 
experience. There are also privacy concerns when audio is continuously 
transmitted to the cloud. To mitigate these concerns, the devices first detect 
predefined keyword(s) such as  "Alexa", "Ok Google", "Hey Siri", etc., which 
is commonly known as keyword spotting (KWS). Detection of keyword wakes up the device 
 and then activates the full scale 
speech recognition either on device~\cite{speech_on_mobile} or in the cloud.
In some applications, the sequence of keywords can be used as voice commands
to a smart device such as a voice-enabled light bulb.
Since KWS system is always-on, it should have very low power consumption to maximize
battery life. 
On the other hand, the KWS system should detect the keywords with 
high accuracy and low latency, for best user experience. 
These conflicting system requirements make KWS an active area of research 
ever since its inception over 50 years ago~\cite{kws_1967}. 
Recently, with the renaissance of artificial neural networks in the form of
deep learning algorithms, neural network (NN) based KWS has become 
very popular~\cite{dnn,cnn,cnn_gru,lstm}.

Low power consumption requirement for keyword spotting systems make 
microcontrollers an obvious choice for deploying KWS in an always-on system. 
Microcontrollers are low-cost energy-efficient processors that are ubiquitous 
in our everyday life with their presence in a variety of devices
ranging from home appliances, automobiles and consumer electronics
to wearables. 
However, deployment of neural network based KWS on microcontrollers 
comes with following challenges:

\textbf{Limited memory footprint}: Typical microcontroller
systems have only tens to few hundred KB of memory available. 
The entire neural network model, including input/output, 
weights and activations, has to fit within this small memory budget.

\textbf{Limited compute resources}: Since KWS is always-on, 
the real-time requirement limits the total number of operations
per neural network inference.

These microcontroller resource constraints in conjunction with the 
high accuracy and low latency requirements of KWS 
call for a resource-constrained neural network architecture exploration
to find {\it lean} neural network structures suitable for KWS, which is
the primary focus of our work.
The main contributions of this work are as follows:
\begin{itemize}
 \item We first train the popular KWS neural net models from the 
literature ~\cite{dnn,cnn,cnn_gru,lstm} on Google speech commands 
dataset~\cite{google_dataset} and compare them
in terms of accuracy, memory footprint and number of operations per inference.
 \item In addition, we implement a new KWS model using depth-wise 
separable convolutions and point-wise convolutions, inspired by the
success of resource-efficient MobileNet~\cite{mobilenet} in computer vision.
This model outperforms the other prior models in all aspects of accuracy,
model size and number of operations. 
 \item Finally, we perform resource-constrained neural network architecture
exploration and present comprehensive comparison of different network 
architectures within a set of compute and memory constraints
of typical microcontrollers. 
The code, model definitions and pretrained models are 
available at \href{https://github.com/ARM-software/ML-KWS-for-MCU}{https://github.com/ARM-software/ML-KWS-for-MCU}.
\end{itemize}

\section{Background}
\subsection{Keyword Spotting (KWS) System}
\begin{figure}[b]
\centering
\includegraphics[width=0.9\textwidth]{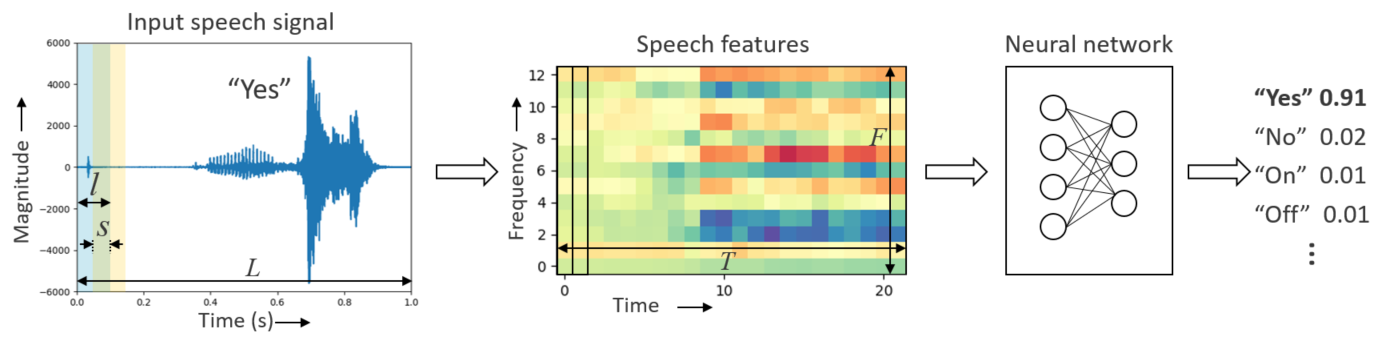}
\caption{\label{fig:mfcc}Keyword spotting pipeline.}
\end{figure}

A typical KWS system consists of a feature extractor and a neural network 
based classifier as shown in Fig.~\ref{fig:mfcc}. First, the input speech 
signal of length $L$ is framed into overlapping frames of length $l$ with 
a stride $s$, giving a total of $T= \frac{L-l}{s}+1$ frames. From each frame, 
$F$ speech features are extracted, generating a total of $T \times F$ features 
for the entire input speech signal of length $L$. Log-mel filter bank energies 
(LFBE) and Mel-frequency cepstral coefficients (MFCC) are the commonly used 
human-engineered speech features in deep learning based speech-recognition, 
that are adapted from traditional speech processing techniques. 
Feature extraction using LFBE or MFCC involves translating the time-domain
speech signal into a set of frequency-domain spectral coefficients, 
which enables dimensionality compression of the input signal.
The extracted speech feature matrix is fed into a classifier module, 
which generates the probabilities for the output classes. In a real-world 
scenario where keywords need to be identified from a continuous audio stream, 
a posterior handling module averages the output probabilities of each output 
class over a period of time, improving the overall confidence of the prediction.

Traditional speech recognition technologies for KWS use Hidden Markov Models 
(HMMs) and Viterbi decoding~\cite{wilpon1990automatic,rose1990hidden}. 
While these approaches achieve reasonable accuracies, they are hard to train and 
are computationally expensive during inference. Other techniques explored 
for KWS include discriminative models adopting a large-margin problem 
formulation~\cite{discriminative_kws} or recurrent neural networks 
(RNN)~\cite{discriminative_rnn}. Although these methods significantly 
outperform HMM based KWS in terms of accuracy, they suffer from large 
detection latency. KWS models using deep neural networks (DNN) based on 
fully-connected layers with rectified linear unit (ReLU) activation 
functions are introduced in~\cite{dnn}, which outperforms the HMM models 
with a very small detection latency. Furthermore, low-rank approximation 
techniques are used to compress the DNN model weights achieving similar 
accuracy with less hardware resources~\cite{tucker2016model,nakkiran2015compressing}. 
The main drawback of DNNs is that they ignore the local temporal and 
spectral correlation in the input speech features. In order to exploit 
these correlations, different variants of convolutional neural network 
(CNN) based KWS are explored in~\cite{cnn}, which demonstrate higher 
accuracy than DNNs. The drawback of CNNs in modeling time varying signals 
(e.g. speech) is that they ignore long term temporal dependencies. 
Combining the strengths of CNNs and RNNs, convolutional recurrent 
neural network based KWS is investigated in~\cite{cnn_gru} and 
demonstrate the robustness of the model to noise. While all the prior 
KWS neural networks are trained with cross entropy loss function, a 
max-pooling based loss function for training KWS model with long 
short-term memory (LSTM) is proposed in~\cite{lstm}, which achieves 
better accuracy than the DNNs and LSTMs trained with cross entropy loss.

Although many neural network models for KWS are presented in 
literature, it is difficult to make a fair comparison between them as 
they are all trained and evaluated on different proprietary datasets 
(e.g.  "TalkType" dataset in~\cite{cnn_gru}, "Alexa" dataset in~\cite{lstm}, 
etc.) with different input speech features and audio duration. Also, 
the primary focus of prior research has been to maximize the accuracy 
with a small memory footprint model, without explicit constraints
of underlying hardware, such as limits on number of operations per inference.
In contrast, this work is more hardware-centric and 
targeted towards neural network architectures that maximize accuracy on 
microcontroller devices. The constraints on memory and compute significantly 
limit the neural network parameters and the number of operations.

\subsection{Microcontroller Systems}\label{sec:mcu_background}

A typical microcontroller system consists of a processor core, an on-chip SRAM
block and an on-chip embedded flash. Table ~\ref{tab:boards} shows some commercially 
available microcontroller development boards with Arm 
Cortex-M processor cores 
with different compute capabilities running at 
different frequencies (16 MHz to 216 MHz), consisting of a wide range of 
on-chip memory (SRAM: 8 KB to 320 KB; Flash: 128 KB to 1 MB). 
The program binary, usually preloaded into the non-volatile flash, 
is loaded into the SRAM at startup and the processor runs the 
program with the SRAM as the main data memory. 
Therefore, the size of the SRAM limits the 
size of memory that the software can use.

\begin{table}[b]
\fontsize{9}{10}\selectfont
\centering
\begin{tabular}{|c|c|c|c|c|}
\hline
Arm Mbed\textsuperscript{TM} platform & Processor & Frequency & SRAM & Flash\\\hline
Mbed LPC11U24 & Cortex-M0 & 48 MHz & 8 KB & 32 KB \\
Nordic nRF51-DK & Cortex-M0 & 16 MHz & 32 KB & 256 KB \\
Mbed LPC1768 & Cortex-M3 & 96 MHz & 32 KB & 512 KB \\
Nucleo F103RB & Cortex-M3 & 72 MHz & 20 KB & 128 KB \\
Nucleo L476RG & Cortex-M4 & 80 MHz & 128 KB & 1 MB \\
Nucleo F411RE & Cortex-M4 & 100 MHz & 128 KB & 512 KB \\
FRDM-K64F & Cortex-M4 & 120 MHz & 256 KB & 1 MB \\
Nucleo F746ZG & Cortex M7 & 216 MHz & 320 KB & 1 MB \\
\hline
\end{tabular}
\vspace{0.2cm}
\caption{\label{tab:boards}Typical off the shelf Arm Cortex-M based 
microcontroller development platforms.}
\end{table}

Other than the memory footprint, performance (i.e., operations per second) is also
a constraining factor for running neural networks on microcontrollers.
Most microcontrollers are designed for embedded applications with low cost
and high energy-efficiency as the primary targets, and do not have
high throughput for compute-intensive workloads such as neural networks.
Some microcontrollers have integrated DSP instructions that can be
useful for running neural network workloads.
For example, Cortex-M4 and Cortex-M7 have integrated SIMD and MAC
instructions that can be used to accelerate low-precision computation
in neural networks. 

\section{Neural Network Architectures for KWS}
This section gives an overview of all the different neural network architectures 
explored in this work including the deep neural network (DNN), convolutional 
neural network (CNN), recurrent neural network (RNN), convolutional recurrent 
neural network (CRNN) and depthwise separable convolutional neural network (DS-CNN).

\subsection{Deep Neural Network (DNN)}
The DNN is a standard feed-forward neural network made of a stack of 
fully-connected layers and non-linear activation layers. The input to the DNN 
is the flattened feature matrix, which feeds into a stack of $d$ hidden 
fully-connected layers each with $n$ neurons. Typically, each fully-connected 
layer is followed by a rectified linear unit (ReLU) based activation function. 
At the output is a linear layer followed by a softmax layer generating the 
output probabilities of the $k$ keywords, which are used for further posterior 
handling.

\subsection{Convolutional Neural Network (CNN)}
One main drawback of DNN based KWS is that they fail to efficiently model the 
local temporal and spectral correlation in the speech features. CNNs exploit 
this correlation by treating the input time-domain and spectral-domain features 
as an image and performing 2-D convolution operations over it. The convolution 
layers are typically followed by batch normalization~\cite{ioffe2015batch}, 
ReLU based activation functions and optional max/average pooling layers, 
which reduce the dimensionality of the features. 
During inference, the parameters of batch normalization can be folded into 
the weights of the convolution layers. In some cases, a linear 
low-rank layer, which is simply a fully-connected layer without non-linear 
activation, is added in between the convolution layers and dense layers for 
the purpose of reducing parameters and accelerating training
~\cite{ba2014deep,sainath2013low}.

\subsection{Recurrent Neural Network (RNN)}
\begin{figure}[b]
\centering
\includegraphics[width=0.85\textwidth]{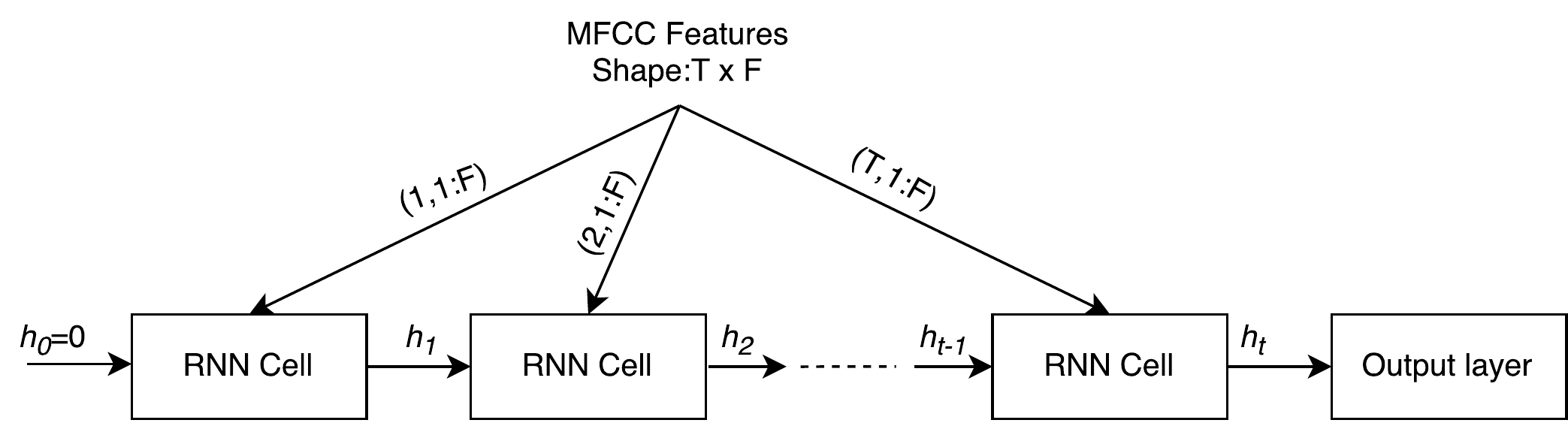}
\caption{\label{fig:RNN}Model architecture of RNN.}
\end{figure}

RNNs have shown superior performance in many sequence modeling tasks, 
especially speech recognition~\cite{sak2014long}, language modeling
~\cite{mikolov2010recurrent}, translation~\cite{sutskever2014sequence}, 
etc. RNNs not only exploit the temporal relation between the input signal, 
but also capture the long-term dependencies using "gating" mechanism. 
Unlike CNNs where input features are treated as 2-D image, RNNs operate 
for $T$ time steps, where at each time step $t$ the corresponding spectral 
feature vector $f_t\in R^{F}$ concatenated with the previous time step 
output $h_{t-1}$ is used as input to the RNN. Figure \ref{fig:RNN} shows 
the model architecture of a typical RNN model, where the RNN cell could be 
an LSTM cell~\cite{hochreiter1997long,gers2002learning} or a gated recurrent 
unit (GRU) cell~\cite{cho2014learning,chung2014empirical}. Since the weights 
are reused across all the $T$ time steps, the RNN models tend to have less 
number of parameters compared to the CNNs. Similar to batch normalization
in CNNs, research show that applying layer normalization can be beneficial 
for training RNNs~\cite{ba2016layer}, in which the hidden states are 
normalized during each time step. 

\subsection{Convolutional Recurrent Neural Network (CRNN)}
\begin{figure}[t]
\centering
\includegraphics[width=0.6\textwidth]{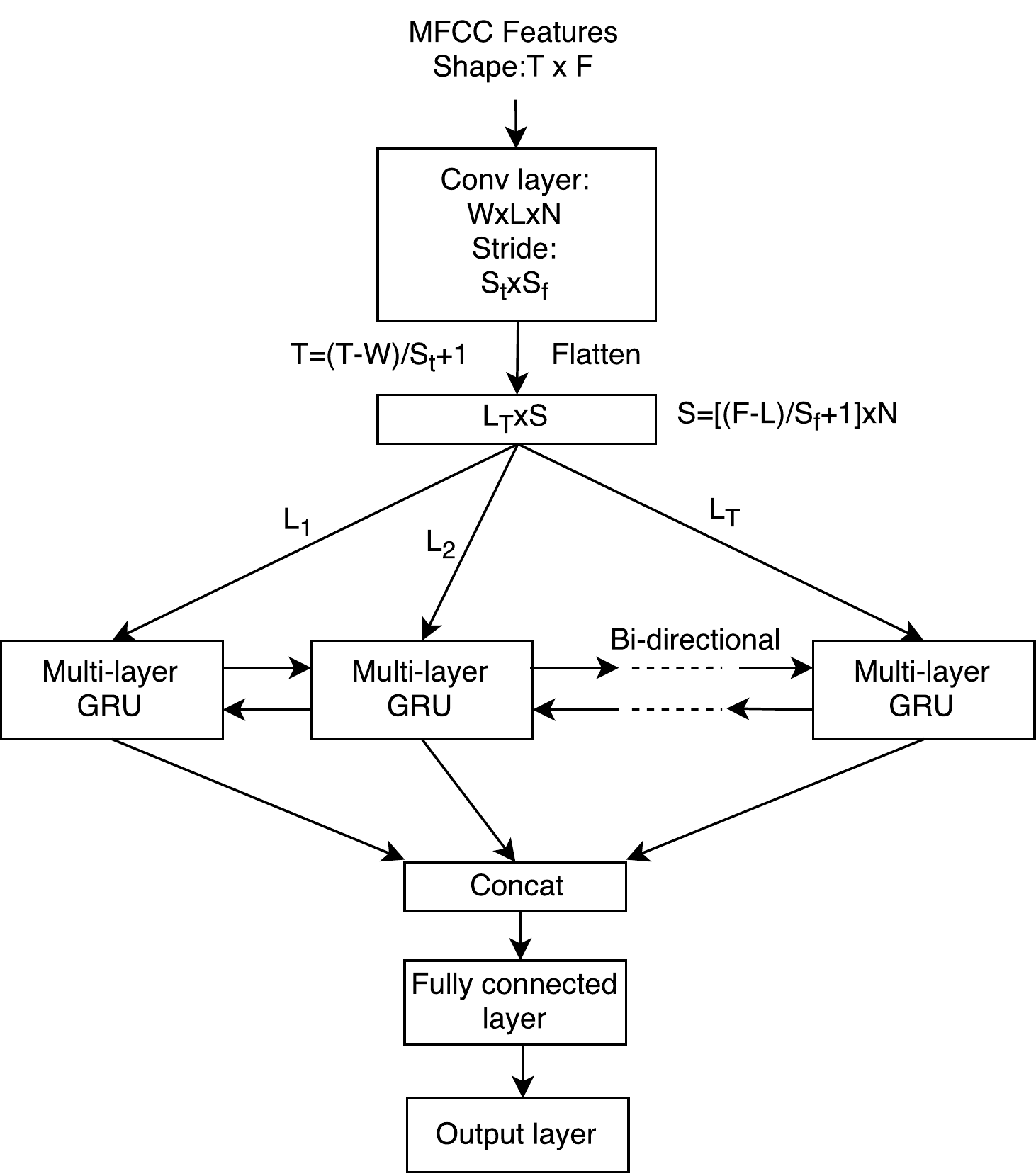}
\caption{\label{fig:CNN-RNN}Model Architecture of CRNN.}
\end{figure}

Convolution recurrent neural network~\cite{cnn_gru} is a hybrid of CNN and RNN, which takes 
advantages of both. It exploits the local temporal/spatial correlation
 using convolution layers and global temporal dependencies in the speech features 
using recurrent layers. As shown in Fig. \ref{fig:CNN-RNN}, a CRNN model starts 
with a convolution layer, followed by an RNN to encode the signal and a dense 
fully-connected layer to map the information. Here, the recurrent layer is 
bi-directional~\cite{schuster1997bidirectional} and has multiple stages, 
increasing the network learning capability. Gated recurrent units (GRU)
~\cite{cho2014learning} is used as the base cell for recurrent layers, 
as it uses fewer parameters than LSTMs and gave better convergence in our 
experiments. 

\subsection{Depthwise Separable Convolutional Neural Network (DS-CNN)}
\begin{figure}[t]
\centering
\includegraphics[width=0.45\textwidth]{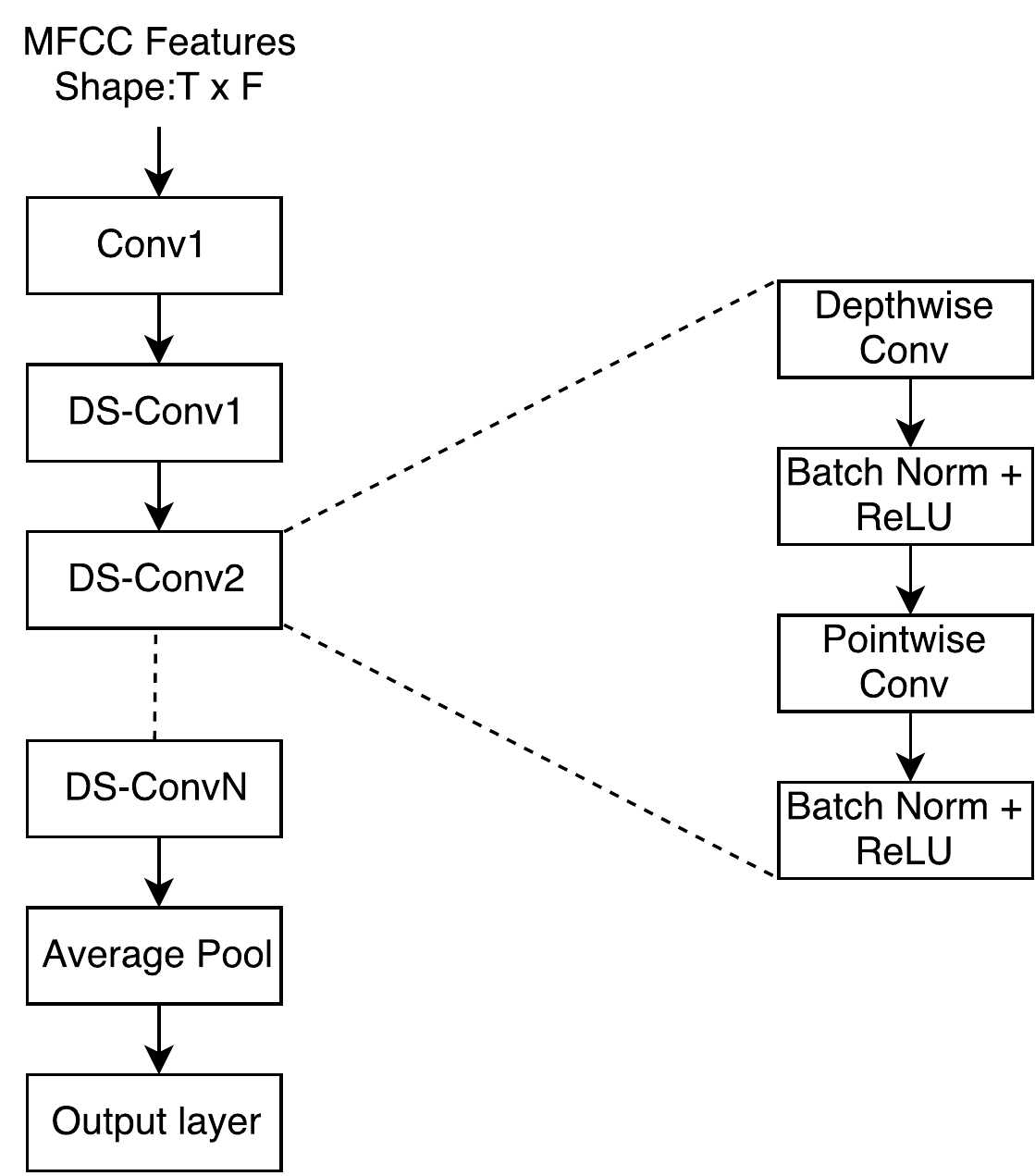}
\caption{\label{fig:mobilenet}Depthwise separable CNN architecture.}
\end{figure}

Recently, depthwise separable convolution has been proposed as an efficient 
alternative to the standard 3-D convolution operation
~\cite{chollet2016xception} and has been used to achieve compact
network architectures in the area of computer vision ~\cite{mobilenet,shufflenet}. 
DS-CNN first convolves each channel in the input feature map 
with a separate 2-D filter and then uses pointwise convolutions (i.e. $1x1$) 
to combine the outputs in the depth dimension. 
By decomposing the standard 3-D convolutions into 2-D convolutions followed
by 1-D convolutions, depthwise separable convolutions are more efficient 
both in number of parameters and operations, which makes deeper and wider 
architecture possible even in the resource-constrained microcontroller devices.
In this work, we adopt a depthwise separable CNN based on the implementation 
of MobileNet~\cite{mobilenet} as shown in Fig. \ref{fig:mobilenet}. An average 
pooling followed by a fully-connected layer is used at the end to provide 
global interaction and reduce the total number of parameters in the final layer.

\section{Experiments and Results}

We use the Google speech commands dataset~\cite{google_dataset} for the
 neural network architecture exploration experiments. The dataset consists 
of 65K 1-second long audio clips of 30 keywords, by thousands of different 
people, with each clip consisting of only one keyword. The neural network 
models are trained to classify the incoming audio into one of the 10 keywords
- "Yes", "No", "Up", "Down", "Left", "Right", "On", "Off", "Stop", "Go",
 along with "silence" (i.e. no word spoken) and "unknown" word, which is the
 remaining 20 keywords from the dataset. The dataset is split into training, 
validation and test set in the ratio of 80:10:10 while making sure that 
the audio clips from the same person stays in the same set. All models are 
trained in Google Tensorflow framework~\cite{abadi2016tensorflow} using the
 standard cross-entropy loss and Adam optimizer~\cite{kingma2014adam}. With 
a batch size of 100, the models are trained for 20K iterations with initial 
learning rate of $5\times10^{-4}$, and reduced to $10^{-4}$ after first 10K 
iterations. The training data is augmented with background noise and random
 time shift of up to 100ms. The trained models are evaluated based on the 
classification accuracy on the test set.

\subsection{Training Results}

Table~\ref{tab:vanilla} summarizes the accuracy, memory requirement and operations
 per inference for the network architectures for KWS from literature
~\cite{dnn,cnn,cnn_gru,lstm} trained on Google speech commands dataset
~\cite{google_dataset}. 
For all the models, we use 40 MFCC features extracted 
from a speech frame of length 40ms with a stride of 20ms, which gives 1960 (49$\times$40) 
features for 1 second of audio. The accuracy shown in the table 
is the accuracy on test set. The memory shown in the table assumes 8-bit 
weights and activations, which is sufficient to achieve same accuracy as that from a 
full-precision network. 
\begin{table}[h]
\fontsize{9}{10}\selectfont
\centering
\begin{tabular}{|c|c|c|c|c|}
\hline
NN Architecture & Accuracy & Memory & Operations \\\hline
DNN~\cite{dnn} &  84.3\%& 288 KB & 0.57 MOps\\
CNN-1~\cite{cnn} & 90.7\% & 556 KB & 76.02 MOps\\
CNN-2~\cite{cnn} &  84.6\%& 149 KB & 1.46 MOps\\
LSTM~\cite{lstm} &  88.8\%& 26 KB & 2.06 MOps\\
CRNN~\cite{cnn_gru} & 87.8\% & 298 KB & 5.85 MOps\\
\hline
\end{tabular}
\vspace{0.2cm}
\caption{\label{tab:vanilla}Neural network model accuracy.
 CNN-1, CNN-2 are (\textit{cnn-trad-fpool3, cnn-one-fstride4})
architectures from~\cite{cnn}.}
\end{table}

Also, we assume that the memory for activations is 
reused across different layers and hence memory requirement for the activations 
uses the maximum of two consecutive layers. The operations in the table counts 
the total number of multiplications and additions in the matrix-multiplication 
operations in each layer in the network, which is representative of the 
execution time of the entire network. The models from the existing literature 
are optimized for different datasets and use different memory/compute resources,
hence a direct comparison of accuracy is unfair. That said, these results still
provide useful insights on the different neural network architectures 
for KWS:

\begin{itemize}
  \item Although DNNs do not achieve the best accuracy and tend to be 
memory intensive, they have less number of operations/inference and hence suit
 well to systems that have limited compute capability (e.g. systems running at 
low operating frequencies for energy-efficiency). 
  \item CNNs, on the other hand, achieve higher accuracy than DNNs 
but at the cost of large number of operations and/or memory requirement. 
  \item LSTMs and CRNNs achieve a balance between memory and operations 
while still achieving good accuracy.
\end{itemize}

\subsection{Classifying Neural Networks for KWS based on Resource Requirements}
\label{sec:classes}

As discussed in section~\ref{sec:mcu_background}, memory footprint and 
execution time are the two important considerations in being able to
run keyword spotting on microcontrollers.
These should be considered when designing and optimizing neural networks
for running keyword spotting.
Based on typical microcontroller system configurations (as described in 
Table~\ref{tab:boards}), we derive three sets of constraints for the
neural networks in Table~\ref{tab:constraints}, targeting small, medium 
and large microcontroller systems.
Both memory and compute limit are derived with assumptions that some
amount of resources will be allocated for running other tasks such as
OS, I/O, network communication, etc.
The operations per inference limit assumes that the system is running
10 inferences per second.

\begin{table}[h]
\fontsize{9}{10}\selectfont
\centering
\begin{tabular}{|c|c|c|}
\hline
NN size & NN memory limit & Ops/inference limit\\\hline
Small (S) & 80 KB & 6 MOps \\
Medium (M) & 200 KB & 20 MOps \\
Large (L) & 500 KB & 80 MOps \\
\hline
\end{tabular}
\vspace{0.2cm}
\caption{\label{tab:constraints}Neural network (NN) classes for KWS models considered 
in this work, assuming 10 inferences per second and 8-bit weights/activations.}
\end{table}
\subsection{Resource Constrained Neural Network Architecture Exploration}

\begin{figure}[h]
\centering
\includegraphics[width=0.8\textwidth]{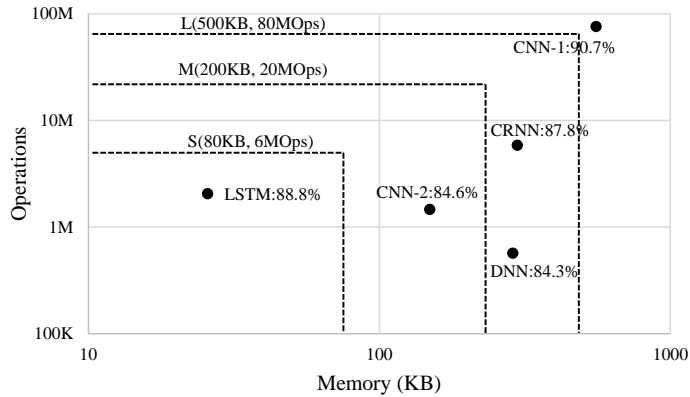}
\vspace{-1.8cm}
\caption{\label{fig:vanilla_bb}Number of operations vs. memory vs. test accuracy 
of NN models from prior work~\cite{dnn,cnn,cnn_gru,lstm} trained on the
speech commands dataset~\cite{google_dataset}.}
\end{figure}

Figure~\ref{fig:vanilla_bb} shows the number of operations per inference, 
memory requirement and test accuracy of neural network models from prior work
~\cite{dnn,cnn,cnn_gru,lstm} trained on Google speech commands dataset 
overlayed with the memory and compute bounding boxes for the neural network 
classes from section~\ref{sec:classes}. 
An ideal model would have high accuracy, small memory footprint and 
lower number of computations, i.e., close to the origin in 
Fig.~\ref{fig:vanilla_bb}. 
Apart from the LSTM model, the other models are too memory/compute resource 
heavy and do not fit into the bounding box \textit{S} with 80KB/6MOps 
memory/compute limits. 
CNN-2, CRNN and DNN models fit in the \textit{M} and \textit{L} bounding 
boxes, but have lower accuracies as compared to the CNN-1 model,
which does not fit in any of the boxes at all. 
The rest of this section discusses different hyperparameters of the
feature extraction and neural network architectures that can be tuned 
in order to bring the models close to the origin and still achieve high accuracy.

\vspace{-1.5cm}

As shown in Fig.~\ref{fig:mfcc}, from each input speech signal, $T \times F$ 
features are extracted and the number of these features impact the model size,
number of operations and accuracy. 
The key parameters in the feature extraction step that impact the model size, number of
operations and accuracy are 
(1) number of MFCC features per frame (\textit{F}) and 
(2) the frame stride (\textit{S}).
The number of MFCC features per audio frame (\textit{F}) impacts the number of 
weights in fully-connected and recurrent layers, but not in convolution 
layers as weights are reused in convolution layers. 
The frame stride (\textit{S}), which determines the number of frames to be 
processed per inference (i.e. \textit{T}), impacts the number of 
weights in fully-connected layers but not in recurrent and convolution layers 
because of the weight reuse. 
Both $F$ and $S$  impact the number of operations per inference. 
An efficient model would maximize accuracy using small $T \times F$, 
i.e., small $F$ and/or large $S$.

The neural network architectures and the corresponding hyperparameters explored
in this work are summarized in Table~\ref{tab:hyperparam}. 
The LSTM model mentioned in the table includes peephole connections 
and output projection layer similar to that in~\cite{lstm}, 
whereas basic LSTM model does not include those.
CRNN uses one convolution layer followed by multi-layer GRU for the recurrent layers. 
We also use batch normalization for convolutional/fully-connected layers and 
layer normalization for recurrent layers. During inference, the parameters of batch 
normalization and layer normalization can be folded into the weights of the 
convolution or recurrent layers and hence these layers are ignored in memory/Ops 
computation.

\begin{table}[h]
\fontsize{9}{10}\selectfont
\centering
\begin{tabular}{|l|l|}
\hline
NN model & Model hyperparameters\\\hline
DNN & Number of fully-connected (FC) layers and size of each FC layer \\
CNN & Number of Conv layers: features/kernel size/stride, linear layer dim., FC layer size \\
Basic LSTM & Number of memory cells \\
LSTM & Number of memory cells, projection layer size \\
GRU & Number of memory cells \\
CRNN & Conv features/kernel size/stride, Number of GRU and memory cells, FC layer size \\
DS-CNN & Number of DS-Conv layers, DS-Conv features/kernel size/stride \\
\hline
\end{tabular}
\vspace{0.2cm}
\caption{\label{tab:hyperparam}Neural network hyperparameters used in this study.}
\end{table}

We iteratively perform exhaustive search of feature extraction hyperparameters and
NN model hyperparameters followed by manual selection to narrow down the search space.
The final best performing models for each neural network architecture along
with their memory requirements and operations are summarized in 
Table~\ref{tab:results_tab} and Fig.~\ref{fig:final_bb}. The hyperparameters of these
networks are summarized in Appendix~\ref{app:hyperparam}.
From the results we can see that DNNs are memory-bound and achieve less accuracies
and saturate at \textasciitilde87\% even when the model is scaled up. 
CNNs achieve better accuracies than DNN, but are limited by the weights 
in the final fully-connected layers. 
RNN models (i.e. Basic LSTM, LSTM and GRU) achieve better accuracies than CNNs and 
yield even smaller models with less Ops in some cases, demonstrating that
exploiting temporal dependencies maximizes accuracy within the same resource budget.
CRNN models, which combine the best properties of CNNs and RNNs, achieve better
accuracies than both CNNs and RNNs, even with less Ops. CRNN architecture also 
scales up well when more memory/compute resources are available.
DS-CNN achieves the best accuracies and demonstrate good scalability owing to 
their deeper architecture enabled by depthwise separable convolution layers, 
which are less compute/memory intensive.

\begin{table*}[htbp]
\fontsize{8}{10}\selectfont
\centering
    \begin{tabular}{|l|c|c|c|c|c|c|c|c|c|}
    \hline
    NN model&\multicolumn{3}{c}{S(80KB, 6MOps)}&\multicolumn{3}{|c|}{M(200KB, 20MOps)}&\multicolumn{3}{c|}{L(500KB, 80MOps)}\\
    \cline{2-10}
    & Acc. & Mem. & Ops & Acc. & Mem. & Ops & Acc. & Mem. & Ops \\ \hline
    DNN	&	84.6\%&  80.0KB&  158.8K&  86.4\%&  199.4KB&  397.0K& 86.7\%&  496.6KB&  990.2K\\  
    CNN	&	91.6\%&  79.0KB&  5.0M&  92.2\%&  199.4KB&  17.3M& 92.7\%&  497.8KB&  25.3M\\  
    Basic LSTM&        92.0\%&  63.3KB&  5.9M&  93.0\%&  196.5KB&  18.9M& 93.4\%&  494.5KB&  47.9M\\  
    LSTM &     92.9\%&  79.5KB&  3.9M&  93.9\%&  198.6KB&  19.2M& 94.8\%&  498.8KB&  48.4M\\ 
    GRU	&	93.5\%&  78.8KB&  3.8M&  94.2\%&  200.0KB&  19.2M& 94.7\%&  499.7KB&  48.4M\\  
    CRNN &	94.0\%&  79.7KB&  3.0M&  94.4\%&  199.8KB&  7.6M& 95.0\%&  499.5KB&  19.3M\\  
    DS-CNN &	94.4\%&  38.6KB&  5.4M&  94.9\%&  189.2KB&  19.8M& 95.4\%&  497.6KB&  56.9M\\  
    \hline
    \end{tabular}
    \vspace{-0.1cm}
    \caption{\label{tab:results_tab}Summary of best neural networks from the hyperparameter search. The memory required for storing the 8-bit weights and activations is shown in the table.}
 \end{table*}

\begin{figure}[h]
\vspace{-2.0cm}
\centering
\subfigure{\includegraphics[clip, trim=2.4cm 4cm 2cm 4cm, width=0.48\columnwidth]{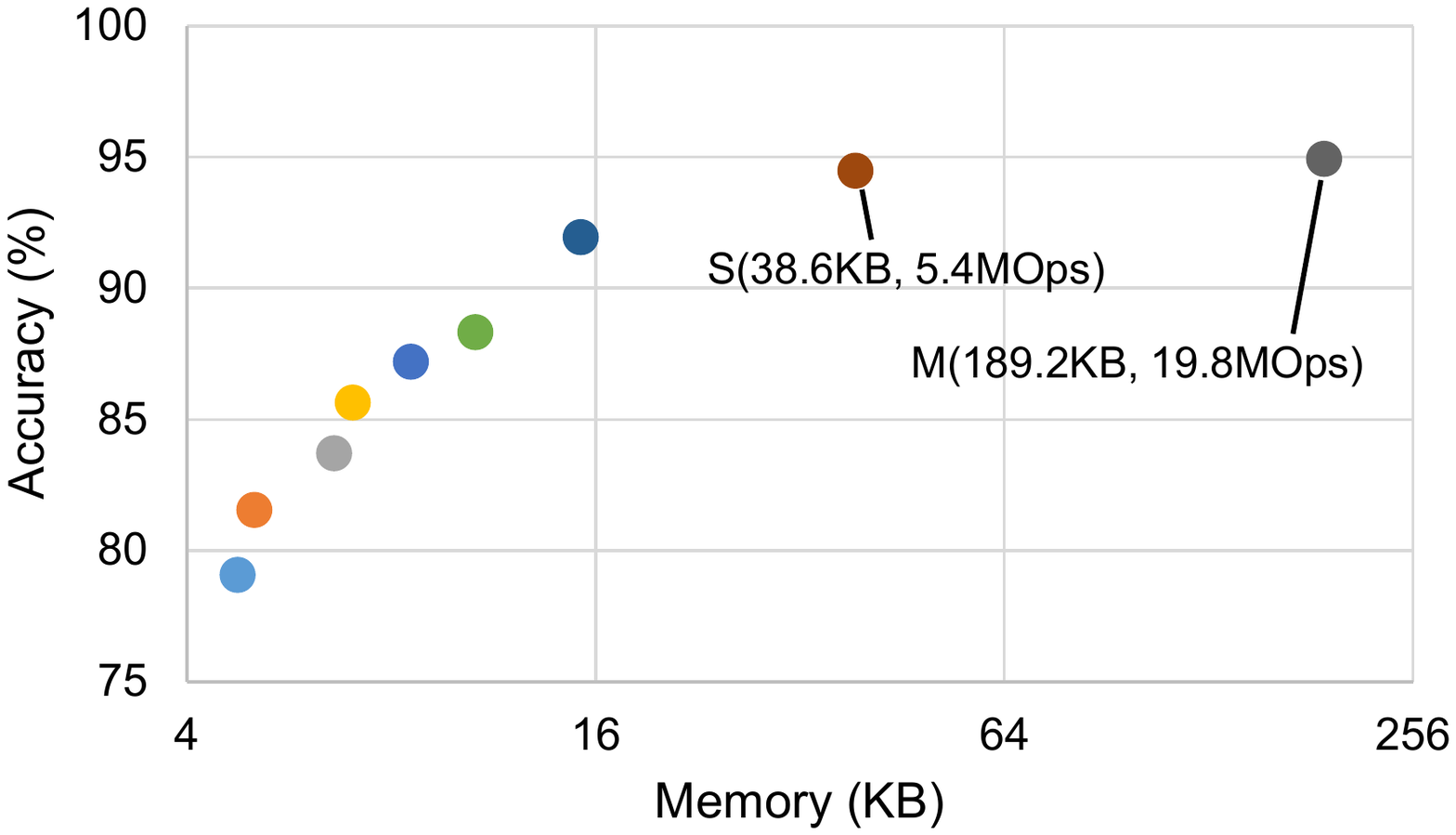}}
\subfigure{\includegraphics[clip, trim=2.4cm 4cm 2cm 4cm, width=0.48\columnwidth]{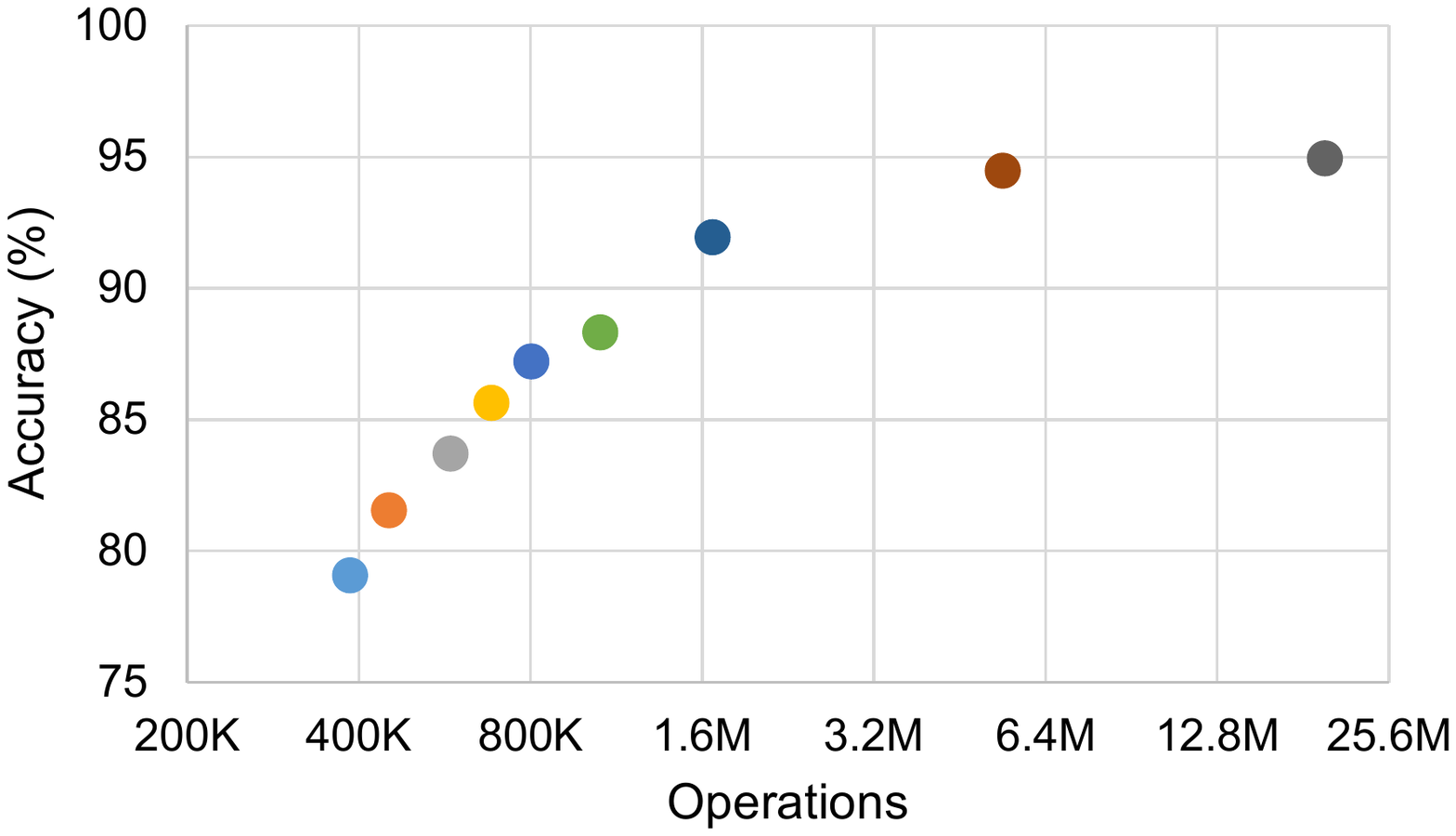}}
\vspace{-2.1cm}
\caption{\label{fig:results_xs}Accuracy vs. memory and Ops of different DS-CNN models demonstrating the scalability of DS-CNN models down to <8KB memory footprint and <500K operations.}
\end{figure}

To study the scalability of the models for smaller microcontroller 
systems with memory as low as 8KB, we expand the search space for DS-CNN models.
Figure~\ref{fig:results_xs} shows the accuracy, memory/Ops requirements of 
the DS-CNN models targeted for such constrained devices. 
It shows that scaled-down DS-CNN models achieve better accuracies than 
DNN models with similar number of Ops, but with >10x reduction in memory requirement.


\begin{figure}[!h]
\vspace{-1.4cm}
\centering
\includegraphics[width=1.0\textwidth]{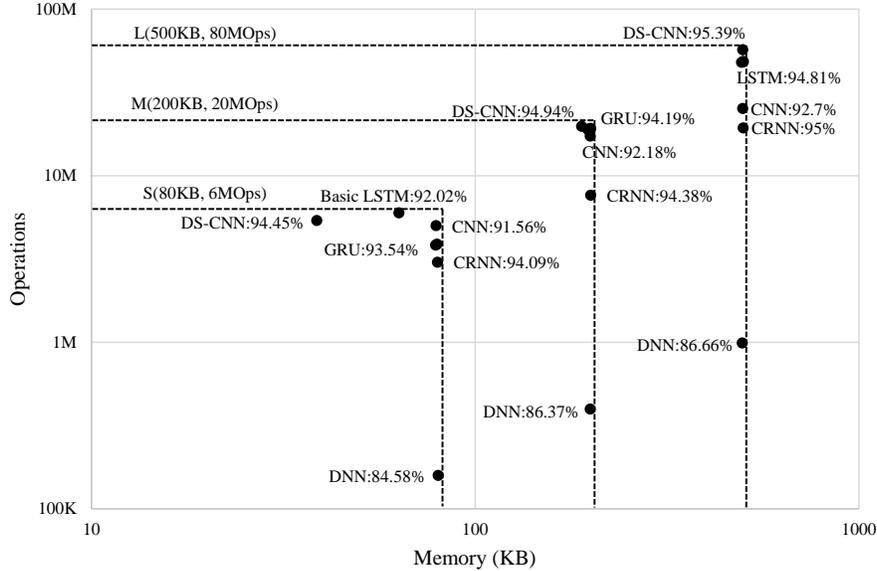}
\vspace{-2.1cm}
\caption{\label{fig:final_bb}Memory vs. Ops/inference of the best models described in Table~\ref{tab:results_tab}.}
\end{figure}

\subsection{Neural Network Quantization}
Neural networks are typically trained with floating point weights and
activations. Previous research~\cite{fpga_paper,dyn_quant,lai2017deep} 
have shown that fixed-point weights is sufficient to run neural 
networks with minimal loss in accuracy. 
Microcontroller systems have limited memory, which
motivates the quantization of 32-bit floating point weights to 8-bit
fixed point weights for deployment, thus reducing the model size by 4$\times$. 
Moreover, fixed-point integer operations run much faster than 
floating point operations in typical microcontrollers, which is another 
reason for executing quantized model during deployment. 

In this work, we use the quantization flow described in~\cite{dyn_quant} 
using 8-bits to represent all the weights and activations. For a given 
signed 2's complement 8-bit fixed-point number, its value ($v$) 
can be expressed as $v=-B_7.2^{7-N}+\sum_{i=0}^{6}B_i.2^{i-N}$, 
where $N$ is the fractional length, 
which can also be negative. $N$ is fixed for a given layer, but can be
different in other layers. For example, 
$N=0$ can represent the range $[-128,127]$ with a step of 1, 
$N=7$ can represent the range $[-1,1-(1/2^7)]$ with a step of $1/2^7$ and 
$N=-2$ can represent the range $[-512,508]$ with a step of $2^2$.

The weights are quantized to 8-bits progressively one layer at 
a time by finding the optimal $N$ for each layer that minimizes the loss in 
accuracy because of quantization. 
After all the weights are quantized, the activations are also quantized 
in a similar way to find the appropriate fractional length $N$ for 
each layer. Table~\ref{tab:quantized_nets} shows the accuracies of 
representative 8-bit networks quantized using this method and 
compared with those of the original full-precision networks. The table
shows that the accuracy of the quantized network is either same or
marginally better than the full-precision network, possibly due to 
better regularization because of quantization. We believe that the 
same conclusion will hold for the other neural network models explored 
in this work. 

\newcolumntype{C}[1]{>{\centering}m{#1}}

\begin{table*}[htbp]
\fontsize{9}{10}\selectfont
\centering
    \begin{tabular}{|l|>{\centering}p{1.2cm}|>{\centering}p{1.2cm}|>{\centering}p{1.2cm}|c|c|c|}
    \hline
    NN model & \multicolumn{3}{c|}{32-bit floating point model accuracy} & 
        \multicolumn{3}{c|}{8-bit quantized model accuracy} \\ 
    \cline{2-7}
    & Train & Val. & Test & Train & Val. & Test \\
    \hline
    DNN	& 97.77\% & 88.04\% & 86.66\% & 97.99\% & 88.91\% & 87.60\%\\ 
    Basic LSTM & 98.38\% & 92.69\% & 93.41\% & 98.21\% & 92.53\% & 93.51\%\\ 
    GRU & 99.23\% & 93.92\% & 94.68\% & 99.21\% & 93.66\% & 94.68\%\\
    CRNN & 98.34\% & 93.99\% & 95.00\% & 98.43\% & 94.08\% & 95.03\%\\
    \hline
    \end{tabular}
    \vspace{0.1cm}
    \caption{\label{tab:quantized_nets}Accuracy comparison of representative
 8-bit quantized networks with full-precision networks.}
 \end{table*}

\subsection{KWS Deployment on Microcontroller}
\begin{figure}[h]
\centering
\includegraphics[width=0.7\textwidth]{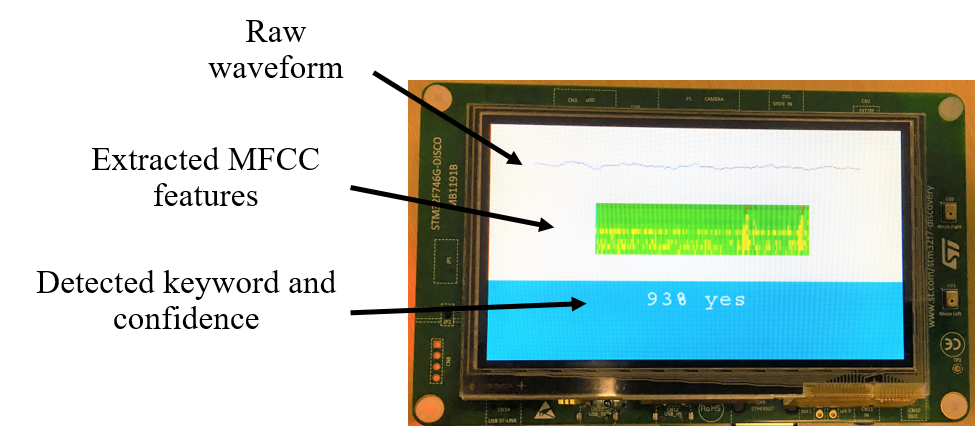}
\caption{\label{fig:kws_demo}
Deployment of KWS on Cortex-M7 development board.
}
\end{figure}

We deployed the KWS application on Cortex-M7 based STM32F746G-DISCO development board
using CMSIS-NN kernels~\cite{lai2018cmsis}.
A picture of the board performing KWS is shown in Fig.~\ref{fig:kws_demo}.
The deployed model is a DNN model with 8-bit weights and 8-bit activations
and KWS is running at 10 inferences per second. Each inference, including memory
copying, MFCC feature extraction and DNN execution, takes about 12 ms. 
The microcontroller can be put into Wait-for-Interrupt (WFI) mode for the 
remaining time for power saving.
The entire KWS application occupies \textasciitilde70 KB memory, including 
\textasciitilde66 KB for weights, \textasciitilde1 KB for activations and 
\textasciitilde~2 KB for audio I/O and MFCC features.

\section{Conclusions}
Hardware optimized neural network architecture is key to get
efficient results on memory and compute constrained microcontrollers.
We trained various neural network architectures for keyword spotting 
published in literature on Google speech commands dataset
to compare their accuracy and memory requirements 
vs. operations per inference, from the perspective of deployment on microcontroller 
systems. 
We quantized representative trained 32-bit floating-point 
KWS models into 8-bit fixed-point versions demonstrating that these models 
can easily be quantized for deployment without any loss in accuracy, 
even without retraining.
Furthermore, we trained a new KWS model using depthwise separable 
convolution layers, inspired from MobileNet.
Based on typical microcontroller systems, we derived three sets of 
memory/compute constraints for the neural networks and performed resource
constrained neural network architecture exploration to find the best networks
achieving maximum accuracy within these constraints. In all three sets of
memory/compute constraints, depthwise separable CNN model (DS-CNN) achieves 
the best accuracies of 94.4\%, 94.9\% and 95.4\% compared to the other 
model architectures within those constraints, which shows good scalability
of the DS-CNN model. The code, model definitions and pretrained models are 
available at \href{https://github.com/ARM-software/ML-KWS-for-MCU}{https://github.com/ARM-software/ML-KWS-for-MCU}.

\section*{Acknowledgements}
We would like to thank Matt Mattina from Arm Research and Ian Bratt from
Arm ML technology group for their help and support.
We would also like to thank Pete Warden from Google's TensorFlow team for his
valuable inputs and feedback on this project.



\newpage
\appendix

\section{Appendix: Neural Network Hyperparameters}\label{app:hyperparam}

Table~\ref{tab:final_hyperparam} shows the summary of the hyperparameters of
the best neural networks described in Table~\ref{tab:results_tab}, along with their
memory, number of operations and accuracy on training, validation and test sets.
All the models use 10 MFCC features, with a frame length (L) of 40ms, where as
the frame stride (S) is shown in the table.
$FC$ stands for fully-connected layer and the number in the parentheses shows the
number of neurons in the fully-connected layer. $C$ stands for convolution layer
and the numbers in parentheses correspond to the number of convolution 
features, kernel sizes in time and frequency axes, strides in time and frequency
axes. Although not shown, all the convolution and fully connected layers have
a ReLU as activation function. $L$ stands for low-rank linear layer with the number
of elements shown in parentheses. 
The number in the parentheses for $LSTM$ and $GRU$ models correspond to the
number of memory elements in those models. 
$DSC$ is depthwise separable convolution layer (DSConv in Fig.~\ref{fig:mobilenet}) 
and the number in the parentheses correspond to the number of features, 
kernel size and stride in both time and frequency axes. 

\begin{table}[h]
\fontsize{9}{10}\selectfont
\centering
\begin{tabular}{|c|c|c|c|c|c|c|c|}
\hline
Model & S & NN model hyperparameters & Memory & Ops & Train & Val. & Test\\
\hline
DNN & 40 & FC(144)-FC(144)-FC(144) & 80.0KB & 158.8K &  91.5\% & 85.6\% & 84.6\% \\
\hline
DNN & 40 & FC(256)-FC(256)-FC(256) & 199.4KB & 397.1K &  95.4\% & 86.7\% & 86.4\% \\
\hline
DNN & 40 & FC(436)-FC(436)-FC(436) & 496.6KB & 990.2K &  97.8\% & 88.0\% & 86.7\% \\
\hline
CNN & 20 & \makecell{C(28,10,4,1,1)-C(30,10,4,2,1)-\\L(16)-FC(128)} & 79.0KB & 5.0M & 96.9\% & 91.1\% & 91.6\% \\
\hline
CNN & 20 & \makecell{C(64,10,4,1,1)-C(48,10,4,2,1)-\\L(16)-FC(128)} & 199.4KB & 17.3M & 98.6\% & 92.2\% & 92.2\% \\
\hline
CNN & 20 & \makecell{C(60,10,4,1,1)-C(76,10,4,2,1)-\\L(58)-FC(128)} & 497.8KB & 25.3M & 99.0\% & 92.4\% & 92.7\% \\
\hline
Basic LSTM & 20 & LSTM(118) & 63.3KB & 5.9M &  98.2\% & 91.5\% & 92.0\% \\
\hline
Basic LSTM & 20 & LSTM(214) & 196.5KB & 18.9M &  98.9\% & 92.0\% & 93.0\% \\
\hline
Basic LSTM & 20 & LSTM(344) & 494.5KB  & 47.9M &  99.1\% & 93.0\% & 93.4\% \\
\hline
LSTM & 40 & LSTM(144), Projection(98) & 79.5KB  & 3.9M &  98.5\% & 92.3\% & 92.9\% \\
\hline
LSTM & 20 & LSTM(280), Projection(130) & 198.6KB  & 19.2M &  98.8\% & 92.9\% & 93.9\% \\
\hline
LSTM & 20 & LSTM(500), Projection(188) & 498.8KB  & 4.8M &  98.9\% & 93.5\% & 94.8\% \\
\hline
GRU & 40 & GRU(154) & 78.8KB & 3.8M &  98.4\% & 92.7\% & 93.5\% \\
\hline
GRU & 20 & GRU(250) & 200.0KB & 19.2M &  98.9\% & 93.6\% & 94.2\% \\
\hline
GRU & 20 & GRU(400) & 499.7KB  & 48.4M &  99.2\% & 93.9\% & 93.7\% \\
\hline
CRNN & 20 & \makecell{C(48,10,4,2,2)-GRU(60)-\\GRU(60)-FC(84)} & 79.8KB & 3.0M & 98.4\% & 93.6\% & 94.1\% \\
\hline
CRNN & 20 & \makecell{C(128,10,4,2,2)-GRU(76)-\\GRU(76)-FC(164)} & 199.8KB & 7.6M & 98.7\% & 93.2\% & 94.4\% \\
\hline
CRNN & 20 & \makecell{C(100,10,4,2,1)-GRU(136)-\\GRU(136)-FC(188)} & 499.5KB & 19.3M & 99.1\% & 94.4\% & 95.0\% \\
\hline
DS-CNN & 20 & \makecell{C(64,10,4,2,2)-DSC(64,3,1)-\\DSC(64,3,1)-DSC(64,3,1)-\\DSC(64,3,1)-AvgPool} & 38.6KB & 5.4M & 98.2\% & 93.6\% & 94.4\% \\
\hline
DS-CNN & 20 & \makecell{C(172,10,4,2,1)-DSC(172,3,2)-\\DSC(172,3,1)-DSC(172,3,1)-\\DSC(172,3,1)-AvgPool} & 189.2KB & 19.8M & 99.3\% & 94.2\% & 94.9\% \\
\hline
DS-CNN & 20 & \makecell{C(276,10,4,2,1)-DSC(276,3,2)-\\DSC(276,3,1)-DSC(276,3,1)-\\DSC(276,3,1)-DSC(276,3,1)-\\AvgPool} & 497.6KB & 56.9M & 99.3\% & 94.3\% & 95.4\% \\
\hline
\end{tabular}
\vspace{0.2cm}
\caption{\label{tab:final_hyperparam}Summary of hyperparameters of the best models described in Table~\ref{tab:results_tab}.}
\end{table}

\begin{figure}[h]
\vspace{-15pt}
\centering
\subfigure[DNN] {
\includegraphics[clip, trim=2cm 9.5cm 2cm 9.5cm, width=0.7\textwidth]{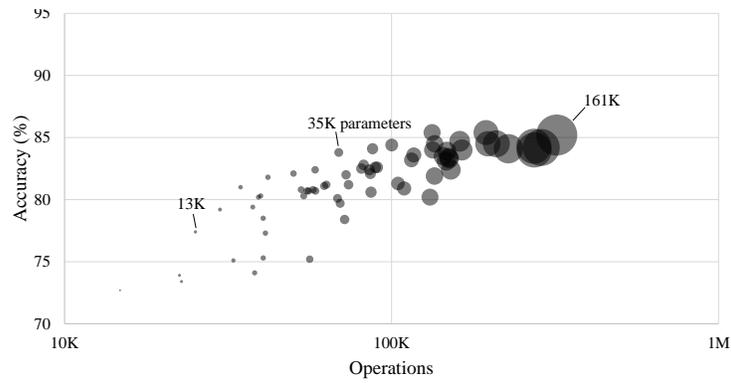}
\label{fig:dnn_hyp}
}
\subfigure[Basic LSTM] {
\includegraphics[clip, trim=2cm 9.5cm 2cm 9.5cm, width=0.7\textwidth]{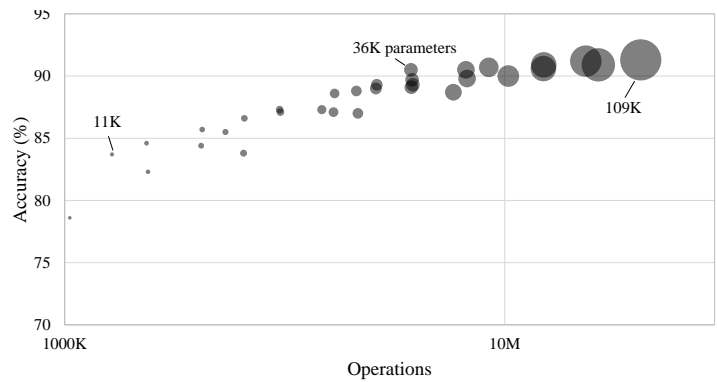}
\label{fig:basic_lstm_hyp}
}
\subfigure[LSTM] {
\includegraphics[clip, trim=2cm 9.5cm 2cm 9.5cm, width=0.7\textwidth]{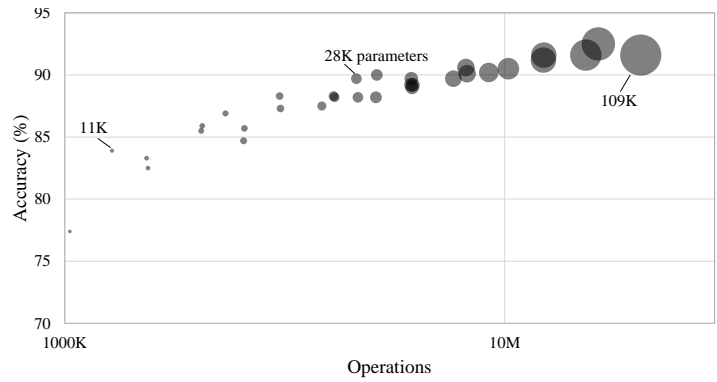}
\label{fig:lstm_hyp}
}
\subfigure[CRNN] {
\includegraphics[clip, trim=2cm 9.5cm 2cm 9.5cm, width=0.7\textwidth]{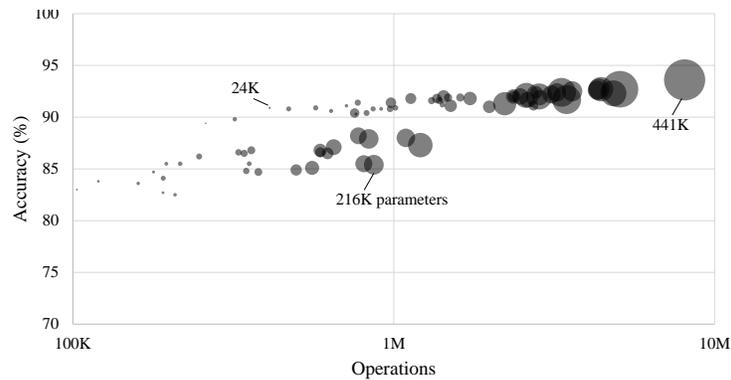}
\label{fig:crnn_hyp}
}
\caption{Hyperparameter search for (a) DNN, (b) basic LSTM,
(c) LSTM and (d) CRNN showing the model accuracy vs. operations, with the number of parameters 
depicted by the size of the circle.}

\end{figure}

Figures~\ref{fig:dnn_hyp},~\ref{fig:basic_lstm_hyp},~\ref{fig:lstm_hyp},~\ref{fig:crnn_hyp} show the 
hyperparameter search of DNN, basic LSTM, LSTM and CRNN architectures depicting the model accuracy vs. 
number of operations. The model size is depicted by the size of the circle.

\end{document}